\magnification=1000\hsize=14truecm\vsize=22truecm
\parindent=4truemm\def\*{\vskip3truemm}\def\\{\hfill\break}
\let\phi=\varphi\let\f=\phi\let\epsilon=\varepsilon\let\x=\xi
\def\fra#1#2{{#1\over#2}}

\centerline{\bf Homoclinic splitting, II. A possible counterexample to a claim}
\centerline{\bf by Rudnev and Wiggins on Physica D.}
\*
\centerline{\it G. Gallavotti, G. Gentile, V. Mastropietro}
\*
\centerline{Universit\`a di Roma 1,2,3 }
\*
{\it Abstract: Results in the mentioned paper do not seem correct.}

\*
{\bf \S1. \it Introduction.}
\*

In the paper [RW] it is claimed that the ``quasi flat''
estimates in the paper [G3] can be improved (see p. 9,17 
of [RW]). We do not understand and below we explain why.

Using the notations of [RW], a point in phase space is $\vec\Gamma=(x,
\vec\phi,y,\vec I)$. In their definition the homoclinic
splitting is the difference in the action coordinates $\vec I$ of the
points $\tilde \Gamma^u$ and $\tilde\Gamma^s$ which are on the stable
and unstable manifolds of an invariant torus and which at time $t=0$
have $x,\vec\phi$ coordinates $x(0,\vec\alpha)=\pi +\xi(\alpha,\mu),
\vec\phi(0,\vec\alpha)=\vec\alpha+\vec\zeta(\vec\alpha,\mu)$ where
$\mu$ is the perturbation parameter. The homoclinic splitting is
$\vec\Delta(\vec\alpha,\mu)= \vec I^s(\vec\alpha)-I^u(\vec\alpha)$.
The $\xi(\vec\alpha,\mu),\vec\zeta(\vec\alpha,\mu)$, ``initial data'',
were not chosen $0$ in [RW], so that $\vec\alpha$ is just a parameter
for the initial data.

We found several points in [RW] that we could not understand and,
thinking that they either were simply too difficult for us or that they
were related to trivial typos, we thought that we could only check the
conclusions in a simple case. We take, first, the model in eq. (8) of
[RW] with a {\it one dimensional} rotator and with $F(x,\phi)=(1-\cos
x)^2\cos\f$:

$$ \omega I+ {y^2\over2}+\epsilon\,(\cos x-1)+\mu(1-\cos
x)^2\cos\f\eqno(1.1)$$
which is a special {\it Thirring model}.  For purposes of comparison
with [G3] model (1.1) translates, after rescaling actions by
$\sqrt\epsilon$ and time by $\epsilon$: $\omega
I\epsilon^{-1/2}+{y^2\over2}+(\cos x-1)+\mu(1-\cos x)^2\cos\f$ where
$\mu$ is now the previous $\mu$ times $\epsilon^{-1}$.  What one has
to prove, as a particular case of theorem 2.1 of [RW], is that the
Fourier transform $\widehat \Delta_k$ of the splitting $\vec
\Delta(\vec\alpha)$ defined in (20) of [RW] verifies the unlabeled
inequality in theorem 2.1 of [RW]:

$$|\widehat \Delta_{k}| \le const\,
e^{-|k|\omega{\pi \over 2\sqrt\epsilon}}\eqno(1.2)$$
where the constant depends upon $k,\mu,\epsilon$ and can be bounded by
a power of $\epsilon$ and of $\mu$.  Hence for $k=2$ it must be, to
leading order as $\epsilon\to0$:

$$|\widehat \Delta_{2}| \le const \, e^{-2\omega {\pi \over
2\sqrt\epsilon}}\eqno(1.3)$$
It is easy to check that this is true at first order.  The second order
is only sketched in [RW], and they seem to claim that it does not
matter; {\it i.e.} that it obviously verifies (1.3) togheter with the
higher orders. What one expects from [G3], {\it unless} the bound
(8.1),(8.2) of [G3] is not optimal, would rather be:

$$|\widehat \Delta^{(2)}_{2}| \le const \, e^{-\omega {\pi \over
2\sqrt\epsilon}}\eqno(1.4)$$
to leading order as $\epsilon\to0$, {\it i.e. much larger}. Since {\it
of course} the same result arises when the number of rotators is $>1$
(it suffices to to think that (1.1) has also a third degree of freedom
whose angle $\f'$ is cyclic) then either there is a cancellation or
there is an inconsistency between [RW] and [G3].

\let\p=\pi
\*
{\bf\S2. \it A second order analysis.}
\*
A cancellation cannot be excluded without a calculation
(or an {\it a priori} argument).  The formalism of [RW] (borrowed
from [G3]) allows easily to perform the calculation. To second
order the splitting vector $\Delta^{(2)}$ is deduced from
eq. (63),(64),(66) of [RW], and one finds:
\def\OO{{\cal O}}
$$\Delta^{(2)}(\alpha)=\int_{-\infty}^{\infty} dt\ \partial_{\phi x}
f(t)\,\OO(\partial_x f)(t)\eqno(2.1) $$
with $\x_i(t)$ defined in eq. (56) of [RW] and:
$\OO(F)(t)=\int_{\rho\infty}^t\big(\x_2(t)\x_1(\tau)-\x_2(\tau)\x_1(t)\big)
F(\tau) d\tau$ if $\rho=\rho(t)$ is the {\it sign} of $t$.  Eq. (2.1),
explictly spelled out {\it up to terms indicated by $+\ldots$ and
contributing to the Fourier transform $\widehat \Delta^{(2)}_k$, for $k=2$,
quantities of order $O(e^{-\pi\fra{\omega}{\sqrt\epsilon}})$ which can
be neglected for our purposes,} becomes (see (1.3)):
$$\eqalign{ &\Delta^{(2)}(\alpha) = -{2}
\int_{-\infty}^\infty
dt\, \rho(t)\cdot(\cos x(\tau) -1)\cdot
\sin x(\tau)  \,\cdot \sin(\alpha+\omega t)\cdot \,
\cr
& \cdot \int_{-\infty}^\infty d\tau\cdot
\big(\x_1(t)\x_2(\tau)-\x_2(t)\x_1(\tau))
\cdot(\cos x(\tau) -1)\cdot\sin x(\tau)\cdot\cos(\alpha+\omega \tau\big)
+\ldots\; , \cr} \eqno(2.2) $$
with $x(\tau)=4\,\hbox{arctg}\, e^{-\sqrt\epsilon \tau}$.
We want to show that $\widehat\Delta^{(2)}_2$ is of order
$O(e^{-\fra1{2\sqrt\epsilon}\pi\omega})$ and not vanishing. The
integrals in (2.2) factorize: they are elementary and can be
successively computed (or found on tables of integrals). We only give
the final result:
$$ \widehat \Delta_2 = \epsilon^{-{5\over2}} \fra{3\p^2\omega^2}{4i}
e^{-\fra12\fra{\pi\omega}{\sqrt\epsilon}}\; ,
\eqno(2.4) $$
to leading order as $\epsilon\to0$.  One can get also the subleading
orders exactly, but there is no point to that since equation (2.4)
contradicts (1.3) hence [RW], while agreeing with (1.4) {\it i.e.}  with
[G3]. We cannot explain this contradiction and, unless it arises from a
misundertsanding by us of the ideas in [RW], it indicates that theorem
2.1 is invalid and theorem 2.3 cannot be deduced from it. We hope that
this note will generate the curiosity of some colleague who will explain
where we err in the above remark, if we do.
\vskip3mm

\noindent{\it Acknowledgements}: We are indebted to G. Benfatto for many
discussions and suggestions.

\vskip3mm

\noindent{}References
\*

[G3] Gallavotti, G.: Reviews on Mathematical Physics, {\bf 6}, 343--
411, 1994.

[RW] Rudnev, M., Wiggins, S.: Physica D, {\bf114}, 3--80, 1998.

\*
\noindent{}Preprints by the Authors in: {\tt http://ipparco.roma1.infn.it}

\end